\documentclass[a4paper,11pt]{article}
\usepackage{amsmath,bm,amssymb,color}

\usepackage{graphicx}
\usepackage{jheppub} 
\usepackage{xspace}

\renewcommand{\tabcolsep}{2em}

\def\gapprox{\lower .7ex\hbox{$\;\stackrel{\textstyle >}{\sim}\;$}}

\newcommand{\NNLOJET}{NNLO\protect\scalebox{0.8}{JET}\xspace}
\allowdisplaybreaks

\begin{document}

\preprint{IPPP/16/39, NSF-KITP-16-067, ZU-TH 18/16}

\title{The NNLO QCD corrections to Z boson production at large transverse momentum}

\author{A.\ Gehrmann--De Ridder$^{a,b,c}$, T.\ Gehrmann$^{b,c}$, E.W.N.\ Glover$^d$, A.\ Huss$^{a}$, T.A.\ Morgan$^d$}

\affiliation{
$^a$Institute for Theoretical Physics, ETH, CH-8093 Z\"urich, Switzerland\\
$^b$Department of Physics, University of Z\"urich, CH-8057 Z\"urich, Switzerland\\
$^c$Kavli Institute for Theoretical Physics, UC Santa Barbara, Santa Barbara, USA \\
$^d$Institute for Particle Physics Phenomenology, Department of Physics, University of Durham, Durham, DH1 3LE, UK}

\abstract{
The transverse momentum distribution of massive neutral vector bosons can be measured to high accuracy at hadron 
colliders. The transverse momentum is caused by a partonic recoil, and is determined by QCD dynamics. 
We compute the single and double-differential transverse momentum distributions for fully inclusive $Z/\gamma^*$ production 
including leptonic decay
 to next-to-next-to-leading order (NNLO) in perturbative QCD. We also compute the same distributions normalised to the cross sections for inclusive $Z/\gamma^*$ production, i.e.\ integrated over the transverse momentum of the lepton pair. 
We compare our predictions for the fiducial cross sections to the 8~TeV data 
set from the ATLAS and CMS collaborations, which both observed a 
tension between data and NLO theory predictions, using the experimental cuts and binning. We find that the inclusion of the NNLO QCD effects does not fully resolve the tension with the data for the unnormalised $p^Z_T$ distribution. However, we observe that normalising the NNLO $Z$-boson transverse momentum distribution by the NNLO Drell--Yan cross section substantially improves the agreement between experimental data and theory, and opens the way for precision QCD studies of this observable.}
\maketitle

\section{Introduction}

The Drell--Yan production of lepton pairs 
mediated through a virtual photon or a $Z$ boson is an important process at hadron colliders such as the Large 
Hadron Collider (LHC). This process has a clean signature with a large event rate, leading to small experimental errors 
over a wide range of energies. It is a key process in probing Standard Model physics, for example fitting parton 
distribution functions (PDFs) and measuring the strong coupling constant $\alpha_s$. Furthermore, it is an important 
background in 
searches for signatures of physics beyond the Standard Model, such as supersymmetry and dark matter. 

There has been a significant amount of effort to ensure that theoretical predictions match the high precision data 
available from the LHC. The inclusive cross section for $Z/\gamma^*$ production is known at NNLO accuracy in 
QCD~\cite{dyNNLO,dynnlo,fewz,vrap}. 
Corrections beyond this order have been studied in the soft-virtual approximation~\cite{dyN3LO}. 
The NNLO QCD corrections have been combined with a resummation of next-to-next-to-leading logarithmic effects 
(NNLL)~\cite{dyresum} which is necessary to predict the transverse momentum distribution of the $Z$ boson at small 
$p_T$ and matched with parton showers~\cite{dyNNLOPS}. The NLO electroweak 
corrections~\cite{dyNLOEW,dyNLOEWPS} and the mixed QCD--EW corrections~\cite{dyQCDEW} have also been 
computed.

\begin{figure}[t]
\centering
\label{fig:pt_recoil}
\includegraphics[width=6.0cm]{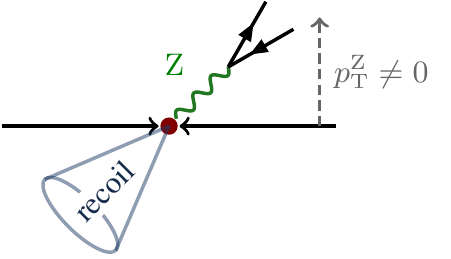}
\caption{A schematic diagram demonstrating the $Z$ boson recoiling against hard radiation.}
\end{figure}

As illustrated in Figure~\ref{fig:pt_recoil}, the transverse momentum of the $Z$ boson is generated by the emission of QCD radiation so that the
 fixed order calculation at ${\cal O}(\alpha_s^2)$, which is NNLO for the inclusive cross section,
  corresponds to only an NLO-accurate prediction of the transverse 
momentum distribution. The NLO EW corrections to the $Z$-boson transverse momentum distribution~\cite{ZJNLOEW} are typically negative and at the level of a few percent at small to moderate $p^Z_T$, but are enhanced by the well-known electroweak Sudakov logarithms when $p^Z_T \gg M_Z$. At small $p^Z_T$, the electroweak corrections approach the corrections to the total $Z$ production cross section, such that they largely cancel when normalising the 
$p^Z_T$ distribution to the total cross section. 

Recently the ${\cal O}(\alpha_s^3)$ NNLO QCD corrections to $Z$+jet production were computed~\cite{ZJNNLOus,ZJNNLOnjet}.
Building upon the calculation presented in Ref.~\cite{ZJNNLOus}, we exploit our highly flexible 
and numerically stable code to compute the transverse momentum distribution 
of the $Z$ boson at finite transverse momentum at NNLO precision. To achieve this we relax the requirement of observing a final state jet and instead impose a low transverse momentum cut on the $Z$ boson. This transverse  momentum cut ensures the infrared finiteness of the NNLO calculation, 
since it enforces the presence of final-state partons 
to balance the transverse momentum of the $Z$ boson.

The production of $Z$ bosons (or, more generally, of lepton pairs with given invariant mass) 
at large transverse momentum has been studied extensively at the LHC by the ATLAS~\cite{ptzATLAS7TeV,ptzATLAS}, CMS~\cite{ptzCMS7TeV,ptzCMS} and LHCb~\cite{ptzLHCb} experiments. 
In order to reduce the systematic 
uncertainty on the measurement, the transverse momentum distribution is commonly normalised to the $p_T$-inclusive 
$Z$-boson production cross section. 
ATLAS and CMS both observed a tension between their measurements and
existing NLO QCD predictions, highlighting the potential importance of higher order corrections to this process. 

Both experiments present their measurements in the form of fiducial cross sections for a restricted kinematical range of 
the final state leptons (in invariant mass, transverse momentum and rapidity). 
In view of a comparison between data and theory, this 
form of presenting the experimental data is preferable over a cross section that is fully inclusive in the lepton kinematics 
(requiring a theory-based extrapolation into phase space regions outside the detector coverage). Consequently, the 
theoretical calculation must take proper account of these restrictions in the final state lepton kinematics.  

The unnormalised $p^Z_T$ distribution represents an absolute cross section measurement based on event counting rates.   
As with any absolute measurement, it has the disadvantage of being sensitive to the proper modelling of 
acceptance corrections, and of relying on the absolute determination of the integrated 
luminosity of the data sample under consideration.  At the LHC the luminosity uncertainty 
alone amounts to about 3\%. In order to reduce the luminosity uncertainty, the data can be normalised to the Drell--Yan cross section for the corresponding fiducial phase space.   This is obtained from the cross section for $Z$ boson production with the same transverse momentum and rapidity cuts on the 
individual leptons, but integrated over all possible transverse momenta of the $Z$ boson.     
On the theoretical side, this amounts to normalising the distribution to the NNLO $pp \to \ell^+\ell^-$+X cross section in which the fiducial cuts are applied to the leptons, but which is fully inclusive on the transverse momentum of the lepton-pair.

In this paper, we compute the NNLO QCD corrections to the transverse momentum distribution for $Z$ production at large transverse momentum, fully inclusive on the accompanying hadronic final state. 
We present the corrections using the final state cuts used by the 
ATLAS~\cite{ptzATLAS} 
and CMS~\cite{ptzCMS} collaborations. 
We study both the absolute (unnormalised) differential cross section and the differential cross section normalised to the relevant Drell--Yan cross section. 
The NNLO corrections lead to a substantially better theoretical description of the experimental data. 
We find that the inclusion of the NNLO QCD effects does not fully resolve the tension with the data for the unnormalised $p^Z_T$ distribution. However, we observe that normalising the NNLO $Z$-boson transverse momentum distribution by the NNLO Drell--Yan cross section leads to an excellent agreement with the experimental distributions.

\section{Setup}
The NNLO corrections to the production of a $Z$ boson at finite transverse momentum
 receive contributions from three types of parton-level processes:  
(a) the two-loop corrections to $Z$-boson-plus-three-parton processes~\cite{Z3p2l}, 
(b) the one-loop  corrections to $Z$-boson-plus-four-parton processes~\cite{Z4p1l,ZJJNLO} and 
(c)  the tree-level $Z$-boson-plus-five-parton  processes~\cite{Z5p0l,ZJJNLO}. 
The ultraviolet renormalised matrix elements for these processes are integrated  over the final state phase space appropriate to 
$Z$ boson final states, including a 
cut on the $p_{T}^{Z}$. All three types of contributions are infrared-divergent and only their sum is finite.  

In this calculation we employ the antenna subtraction method~\cite{ourant} to isolate the infrared singularities 
in the different contributions to enable their cancellation prior to the numerical implementation. 
The construction of the calculation and the subtraction terms is exactly as described in Ref.~\cite{ZJNNLOus}. 
All partonic channels relevant to lepton pair production through the exchange of 
an on-shell or off-shell $Z$ boson or a virtual photon are included. Our calculation is 
implemented in a newly developed parton-level Monte Carlo generator \NNLOJET. This program
provides the necessary 
infrastructure for the antenna subtraction of hadron collider processes at NNLO and performs the integration 
of all contributing subprocesses at this order. Components of it have also been used in 
other NNLO QCD calculations~\cite{eerad3,nnlo2j,nnlott,nnlohj} using the antenna subtraction method. 
Other processes can be added to \NNLOJET provided the matrix elements are 
available. 

To describe the normalised distributions, we also implemented 
the NNLO QCD corrections to inclusive $Z/\gamma^*$ production including leptonic decays in \NNLOJET, and validated this 
implementation against the publicly available codes FEWZ~\cite{fewz} and Vrap~\cite{vrap}. 

For our numerical computations, we use the NNPDF3.0 parton distribution functions (PDFs)~\cite{nnpdf}
with the value of $\alpha_s(M_Z)=0.118$ at NNLO, and $M_Z=91.1876~$GeV. Note that we systematically use the 
same set of PDFs and the same value of $\alpha_s(M_Z)$ for the NLO and NNLO predictions. The factorisation and renormalisation scales are chosen dynamically on an event-by-event basis as,
\begin{equation}
  \label{eq:scale}
\mu \equiv \mu_R = \mu_F = \sqrt{m^2_{\ell\ell} + (p^Z_T)^2},
\end{equation}
where $m_{\ell\ell}$ and $p^Z_T$ are the invariant mass and the transverse momentum of the final state lepton pair respectively. The theoretical uncertainty is estimated by varying the scale choice by a factor in the range $[1/2,2]$. 
The electroweak coupling constant $\alpha$ is derived from the Fermi constant in the $G_\mu$ scheme, which absorbs large logarithms of the light fermion masses induced by the running of the coupling constant from the Thomson limit ($Q^2=0$) to the electroweak scale into the tree-level coupling.
We also impose a cut on the transverse momentum of the $Z$ boson, $p_T^Z > 20$~GeV. In the low transverse momentum region, large logarithmic 
corrections of the form $\log^n(p_T/M_Z)$ appear at each order in the perturbative expansion in $\alpha_s$, spoiling 
its convergence. A reliable theoretical prediction in this region can only be obtained by resummation~\cite{dyresum}  of 
these logarithms to all orders in perturbation theory. The cut on the transverse momentum also ensures the applicability of our 
approach, originally developed for $Z$+jet production at NNLO. In the computation of the inclusive lepton pair production cross section used to normalise the transverse momentum distributions, we choose the same scale~\eqref{eq:scale} but varied independently over the same range of scale variation. The inclusive cross section in this case is, however, dominated by the regime where $\mu^2 \approx m^2_{\ell\ell}$.

\section{The {\boldmath$Z$} boson transverse momentum distribution}

The experimental measurements of $Z$-boson production at finite transverse momentum are presented in the 
form of fiducial cross sections over a restricted phase space for the final state leptons, which is fully contained in 
the detector's coverage. The NNLO corrections to the transverse momentum distribution can be compared to data by considering the same cuts to the lepton kinematics as presented in the ATLAS~\cite{ptzATLAS} and CMS~\cite{ptzCMS}
analyses using data from Run 1 of the LHC with $\sqrt{s}=8$~TeV, which are summarised in Table~\ref{tab:fiducial}. 
In this section we will focus on the ATLAS measurement in the fiducial region defined by a broad dilepton invariant mass 
window around the $Z$ resonance, $66~\text{GeV} < m_{\ell\ell} < 116~\text{GeV}$, and compare both the absolute and normalised $p_T^Z$ distribution to the experimental data.
\begin{table}
\begin{center}
\begin{tabular}{ccc}
& ATLAS & CMS\\ \hline
leading lepton  & $|\eta_{\ell_1}|<2.4$ &  $|\eta_{\ell_1}|<2.1$ \\
& $p^{\ell_1}_{T} > 20$~GeV &  $p^{\ell_1}_{T} > 25$~GeV \\
 sub-leading lepton & $|\eta_{\ell_2}|<2.4$ &  $|\eta_{\ell_2}|<2.4$ \\
 & $p^{\ell_2}_{T} > 20$~GeV &  $p^{\ell_2}_{T,2} > 10$~GeV \\
\hline
\end{tabular}
\end{center}
\caption{Kinematical cuts used to define the fiducial phase space for the final state leptons in the 
measurements of ATLAS~\protect{\cite{ptzATLAS}} and CMS~\protect{\cite{ptzCMS}}.\label{tab:fiducial}}
\end{table}

\begin{figure}
  \centering
\includegraphics[width=0.5\textwidth]{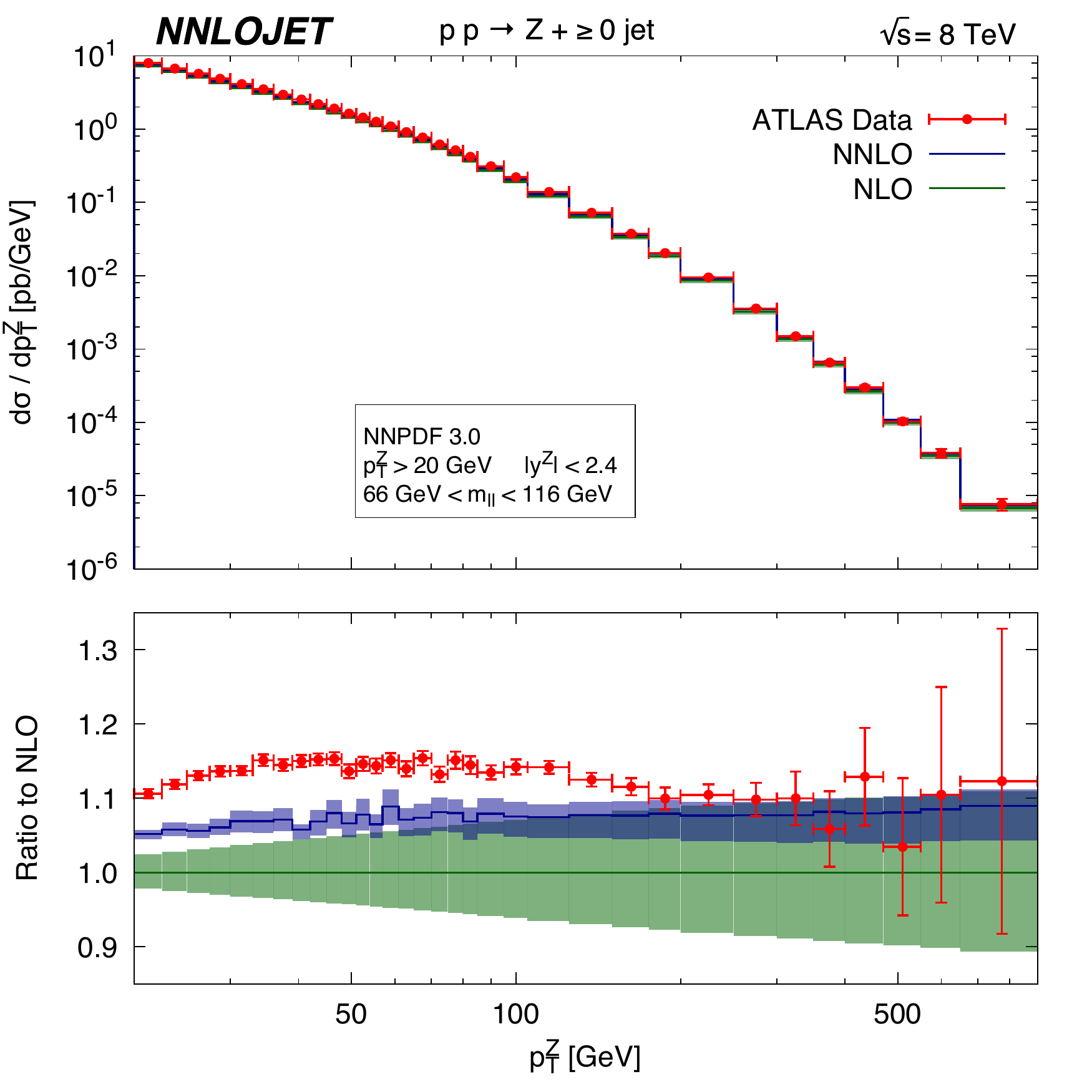}
\caption{The unnormalised $Z$-boson  transverse momentum distribution for the cuts given in Table~\ref{tab:fiducial} and $66~\text{GeV} < m_{\ell\ell} < 116~\text{GeV}$. ATLAS data is taken from Ref.~\cite{ptzATLAS}. The luminosity error is not shown. The green bands denote the NLO prediction with scale uncertainty and the blue bands show the NNLO prediction with scale uncertainty. \label{fig:unnormdsigdpt}}
\end{figure}

Figure~\ref{fig:unnormdsigdpt} shows a comparison of data from the ATLAS analysis within this fiducial region. At low transverse momentum, the NNLO correction increases the cross section by about 6\% (compared to NLO) and significantly reduces the scale uncertainty.  However, there is still significant tension between the ATLAS data and the NNLO prediction.
\begin{figure}
  \centering
\includegraphics[width=0.5\textwidth]{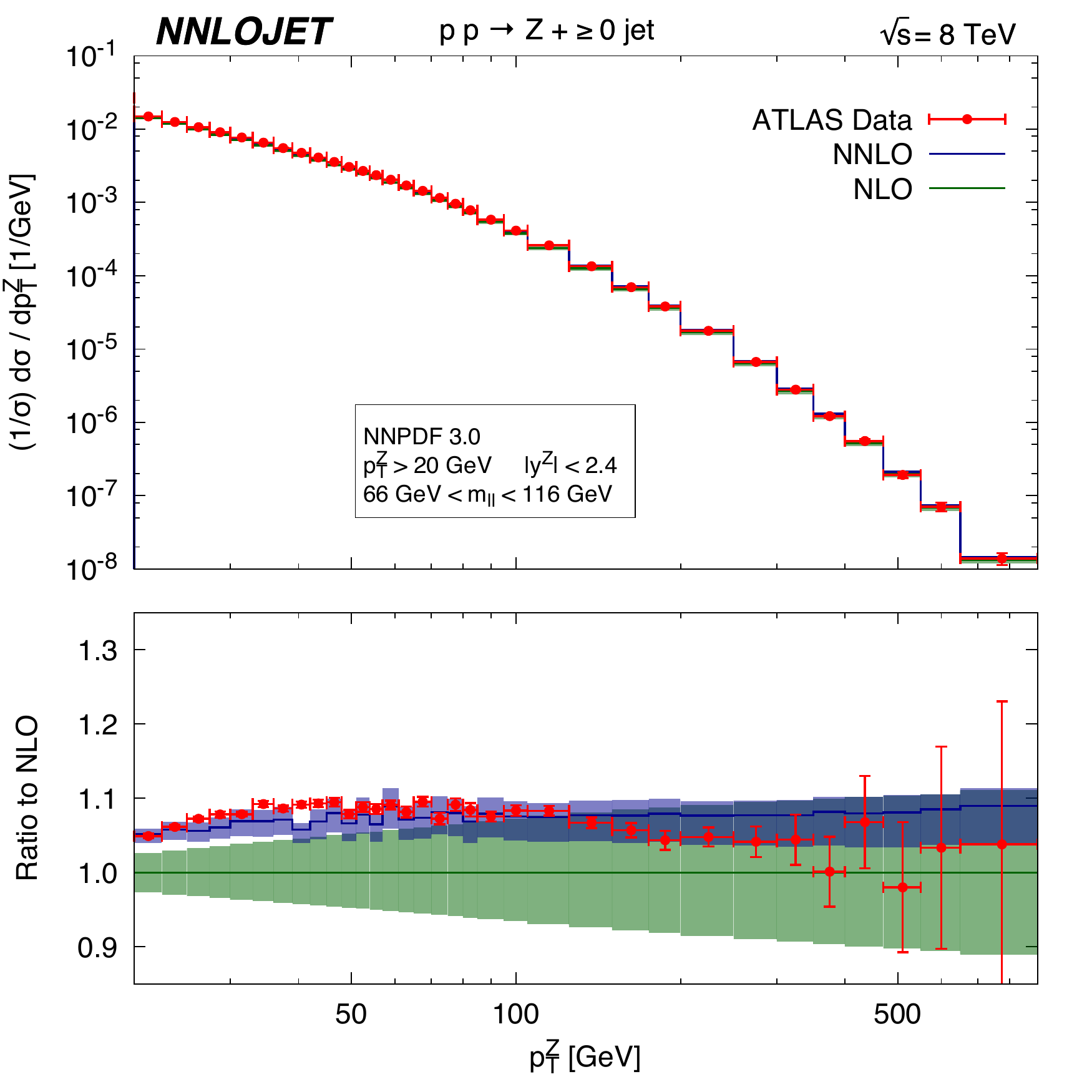}
\caption{The normalised $Z$-boson transverse momentum distribution for the cuts given in Table~\ref{tab:fiducial}  and $66~\text{GeV} < m_{\ell\ell} < 116~\text{GeV}$. ATLAS data is taken from Ref.~\cite{ptzATLAS}. The green bands denote the NLO prediction with scale uncertainty and the blue bands show the NNLO prediction with scale uncertainty. \label{fig:normdsigdpt}}
\end{figure}

The data presented in Figure~\ref{fig:unnormdsigdpt} does not include the error on the integrated luminosity, which amounts to 
an overall normalisation uncertainty of 2.8\% on all data points. 
To cancel the luminosity uncertainty from the measured data, it is more appropriate to normalise 
the data by the measured values for the inclusive 
lepton pair cross section in this fiducial bin. 
The cross section for this mass window was measured to be~\cite{ptzATLAS},
\begin{equation*}
\sigma_{\text{exp}}(66~\text{GeV} < m_{\ell\ell} < 116~\text{GeV}) =  537.10 \pm 0.45\% \text{ (sys.)} \pm 2.80\% \text{ (lumi.)} \text{ pb}.
\end{equation*}
For the NNLO prediction of the cross section in this fiducial region, we obtain,
\begin{equation*}
\sigma_{\text{NNLO}}(66~\text{GeV} < m_{\ell\ell} < 116~\text{GeV}) =  507.9^{+2.4}_{-0.7} \text{ pb}.
\end{equation*}
We see that there is some tension between the measured cross section compared to the theoretical result.  
The normalised distribution is presented in Figure~\ref{fig:normdsigdpt}, where excellent agreement is observed between the normalised NNLO prediction and the experimental data across a wide range of transverse momentum.
The scale bands on the normalised theory predictions are obtained by independently varying the scale in the numerator and denominator, where we use the NNLO prediction for the inclusive Drell--Yan cross section in the normalisation throughout.

The electroweak corrections to this distribution are known to be small at moderate transverse momenta but become more sizeable in the high-$p_T^Z$ tail where they can reach $\sim-15\%$ for $p_T^Z\gtrsim 600$~GeV~\cite{ZJNLOEW}.
It can be therefore expected that the the inclusion of electroweak corrections will further improve the prediction of the shape in the tail.
However, this region of the distribution is still dominated by the statistical uncertainties of the experimental data and a fully consistent inclusion of electroweak corrections is beyond the scope of this work.

\section{Double-differential distributions}

We now compare our theoretical predictions for dilepton production at large transverse momentum for different ranges of  (a) the dilepton mass and (b) the dilepton rapidity.   These double-differential distributions have also been studied by  
ATLAS~\cite{ptzATLAS} and CMS~\cite{ptzCMS} 
with Run 1 data 
using the lepton rapidity and transverse momentum cuts  summarised in Table~\ref{tab:fiducial}.

\begin{figure}
  \centering
\includegraphics[width=\textwidth]{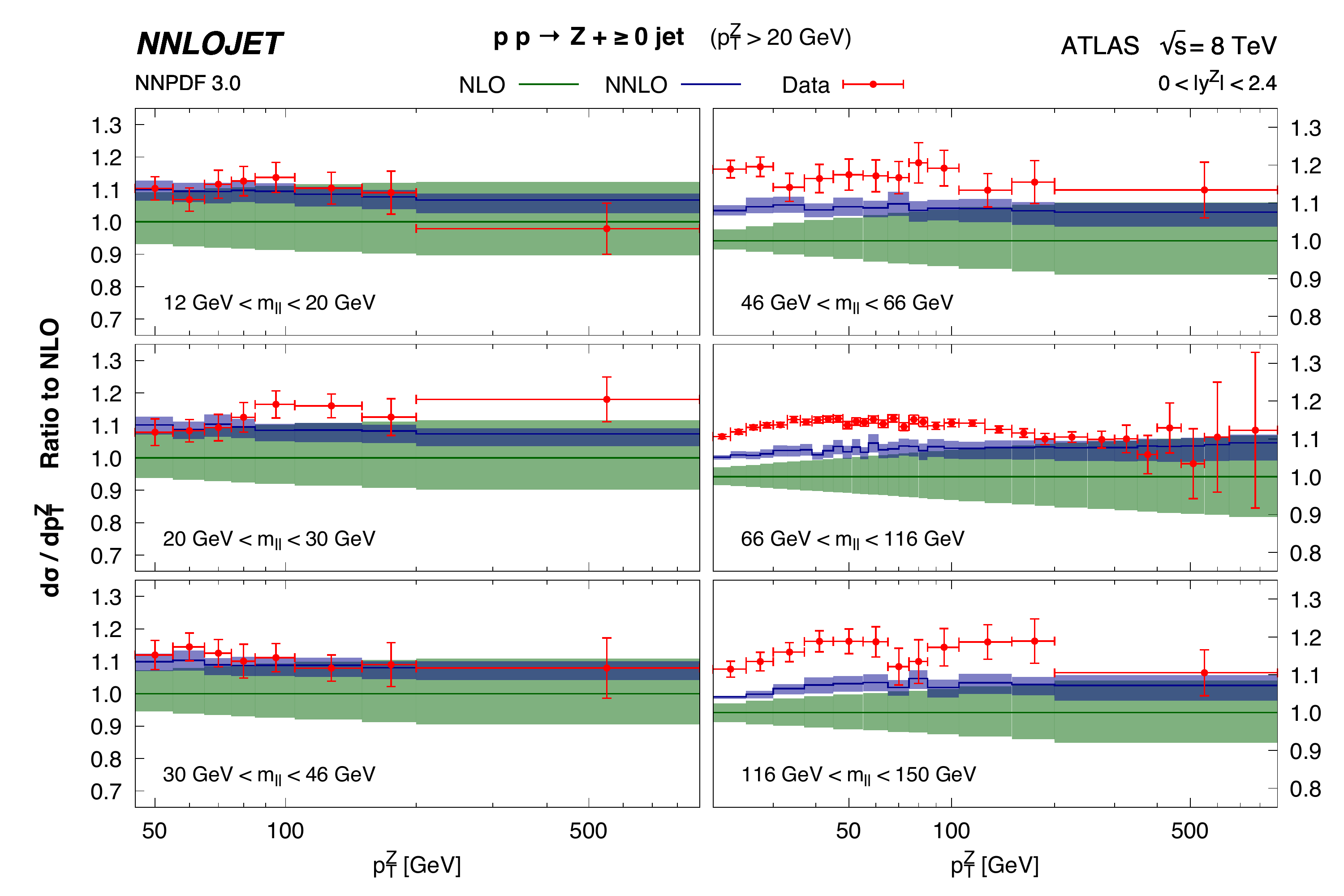}
\caption{The unnormalised double-differential transverse momentum distribution for the $Z$ boson in windows of invariant mass of the leptons, $m_{\ell\ell}$, with a rapidity cut on the $Z$ boson of $|y^Z| < 2.4$. The ATLAS data is taken from Ref.~\cite{ptzATLAS}. The luminosity error is not shown. The green bands denote the NLO prediction with scale uncertainty and the blue bands show the NNLO prediction with scale uncertainty. \label{fig:unnormATLASmll}}
\end{figure}
We first consider the ATLAS measurement, which covers a broader kinematical range. 
In Figure~\ref{fig:unnormATLASmll} we present the unnormalised double-differential distribution with respect to the transverse 
momentum of the $Z$ boson and the invariant mass of the lepton pair, $m_{\ell\ell}$, normalised to the NLO prediction and
compare it to the ATLAS data from Ref.~\cite{ptzATLAS}.
We observe that there is 
good agreement between the NNLO prediction and the data in the low $m_{\ell\ell}$ mass windows. 
For values of $m_{\ell\ell}$ close to the $Z$ boson mass, where the statistical accuracy of the data is highest, 
the NNLO prediction is below the data by about 5-8\%.

\begin{figure}
  \centering
\includegraphics[width=0.6\textwidth]{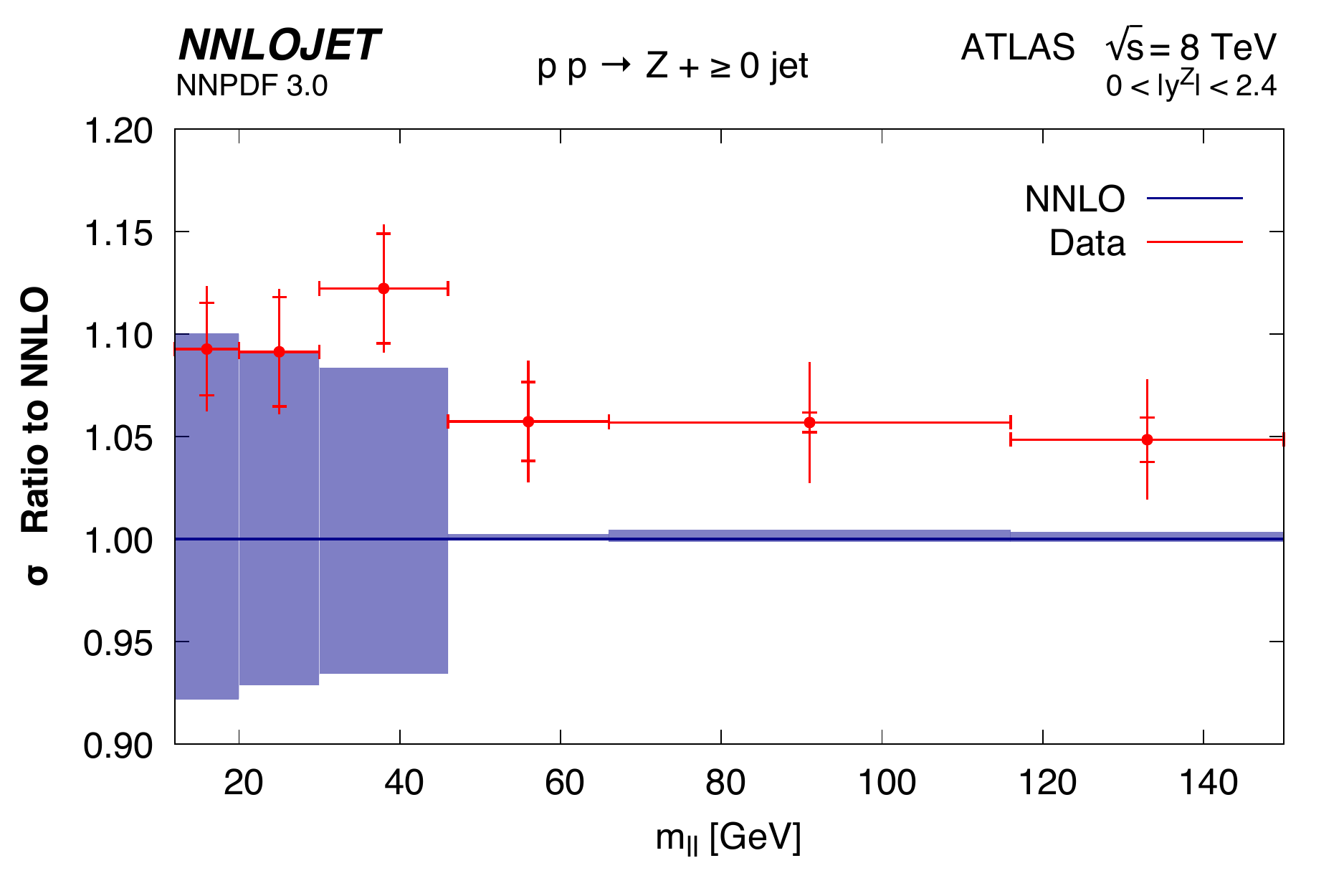}
\caption{The inclusive dilepton cross section for the same $m_{\ell\ell}$ bins as in Figure~\ref{fig:unnormATLASmll} and with a rapidity cut on the $Z$ boson of $|y^Z| < 2.4$. The experimental data is taken from the ATLAS analysis in Ref.~\cite{ptzATLAS}. The ticks on the vertical error bands denote the systematic uncertainty from the measurement, the vertical bars without the ticks are the luminosity uncertainty only. The blue bands show the NNLO prediction with scale uncertainty. \label{fig:incATLASmll}}
\end{figure}
To obtain normalised distributions, these data sets are divided by the inclusive dilepton cross sections for each
fiducial bin, defined by the lepton cuts given in Table~\ref{tab:fiducial} and the appropriate dilepton invariant mass cut. We computed 
these fiducial cross sections to NNLO accuracy (${\cal O}(\alpha_s^2)$)
and compare them to the measured values quoted by ATLAS~\cite{ptzATLAS}
in Figure~\ref{fig:incATLASmll}. We clearly observe that the central value of the NNLO prediction falls below the experimental 
data in all mass bins. The large scale dependence in the three lowest mass bins can be explained from the 
interplay between the cuts on the lepton pair invariant mass and on the single lepton transverse momentum. The latter cuts
forbid events at low transverse momentum of the lepton pair down to $m_{\ell\ell}=40$~GeV, such that the two lowest bins 
receive no leading order contribution, and the third bin only a very small one. All three bins are populated only from NLO 
onwards, with events containing a low-mass lepton pair recoiling against a parton at high transverse momentum. So our NNLO 
prediction for the inclusive cross section in these mass bins is effectively only NLO accurate, with consequently larger 
scale dependence. In the three bins with larger $m_{\ell\ell}$, the scale uncertainty on the NNLO prediction is below 
0.7\%, which 
results in tension between data and theory at the level of two standard deviations.

\begin{figure}
  \centering
\includegraphics[width=\textwidth]{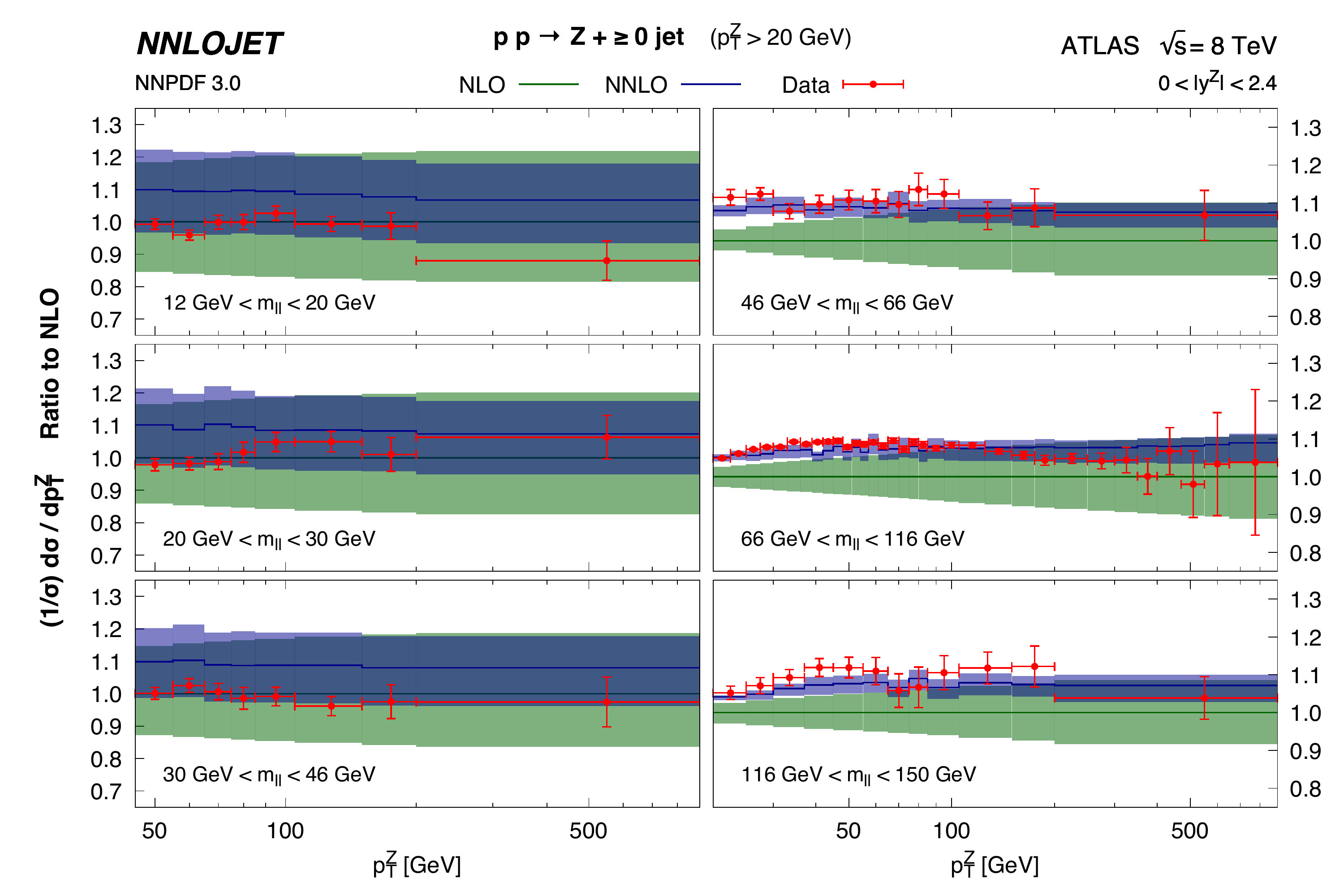}
\caption{The normalised double-differential transverse momentum distribution for the $Z$ boson in windows of invariant mass of the leptons, $m_{\ell\ell}$, with a rapidity cut on the $Z$ boson of $|y^Z| < 2.4$. The ATLAS data is taken from Ref.~\cite{ptzATLAS}.  The green bands denote the NLO prediction with scale uncertainty and the blue bands show the NNLO prediction with scale uncertainty. \label{fig:normATLASmll}}
\end{figure}
Combining together the unnormalised differential distribution with the inclusive cross sections, we obtain the normalised distributions shown in Figure~\ref{fig:normATLASmll}.
Because of the large scale uncertainty in the inclusive cross section, the theoretical errors dominate the low $m_{\ell\ell}$ bins.  At large $m_{\ell\ell}$, the tension between the data and NNLO theory is largely relieved. At the highest values of $p_T^Z$, the tendency of the data to fall below the theory prediction may be an indication 
of the onset of electroweak corrections~\cite{ZJNLOEW}, which are negative in this region.  
Any remaining tension for medium values of $p_T^Z$ could potentially be accounted for revisiting the parton distribution functions 
(especially the gluon distribution) in the kinematical region relevant to this measurement.

\begin{figure}
  \centering
\includegraphics[width=\textwidth]{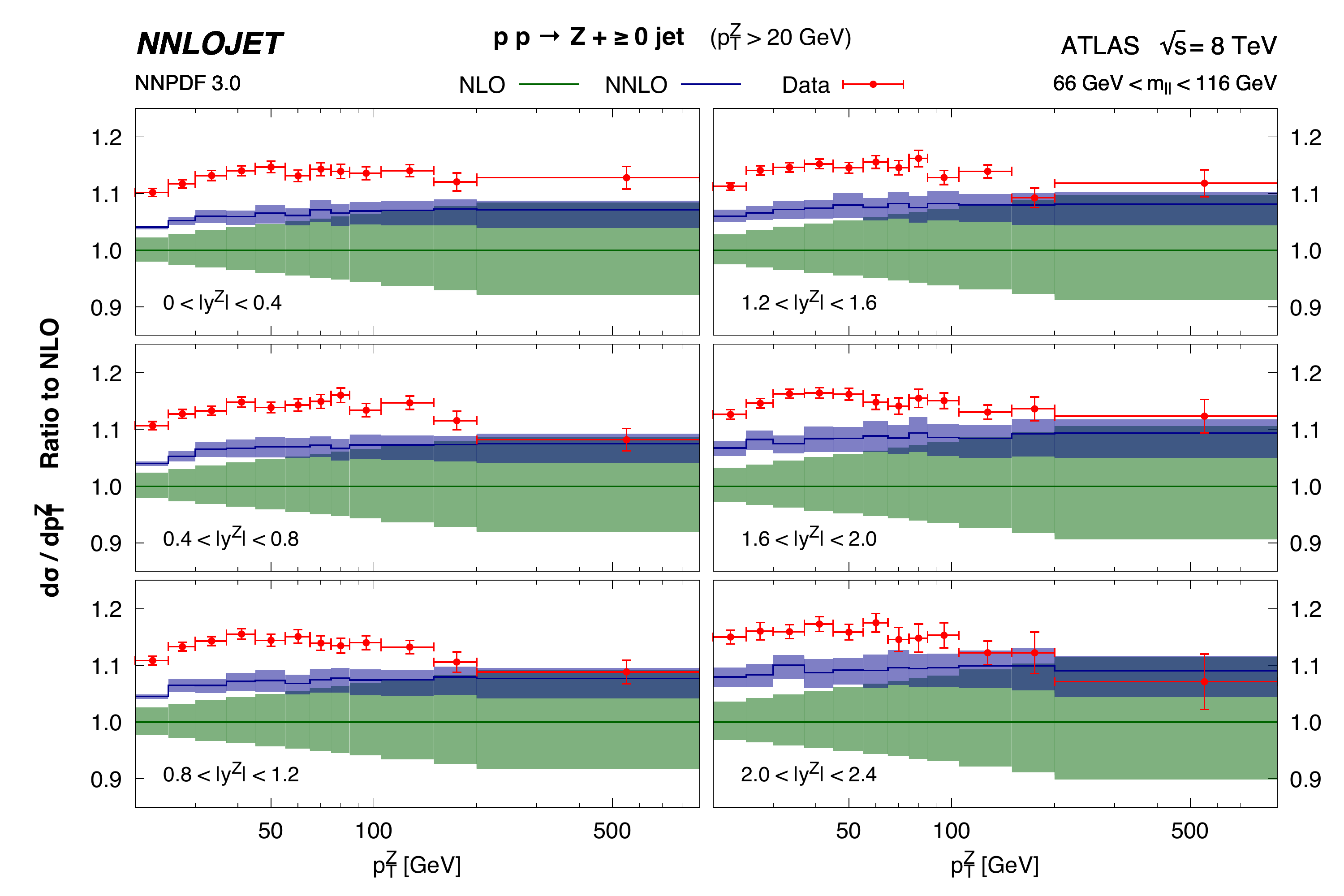}
\caption{The unnormalised double-differential transverse momentum distribution for the $Z$ boson in windows of rapidity of the $Z$ boson, $y^Z$, with an invariant mass cut on the final state leptons of $66$~GeV~$<m_{\ell\ell} < 116$~GeV. The ATLAS data is taken from Ref.~\cite{ptzATLAS}. The luminosity error is not shown. The green bands denote the NLO prediction with scale uncertainty and the blue bands show the NNLO prediction.\label{fig:ATLASyz}}
\end{figure}

The same tension between NNLO theory and ATLAS data for the unnormalised distribution is visible in 
Figure~\ref{fig:ATLASyz}, which 
 shows the unnormalised double-differential distribution with respect to the transverse momentum 
of the $Z$ boson for $66$~GeV~$<m_{\ell\ell} < 116$~GeV in 
 different ranges of the rapidity of the $Z$ boson, normalised to the NLO prediction. The NNLO corrections are uniform 
 in rapidity and transverse momentum at the level of about 5--8\%, and they decrease the 
 residual theoretical uncertainty to 2--4\%. The data, which correspond to the invariant mass bin containing the 
 $Z$ resonance, are consistently above the NNLO prediction. 
\begin{figure}
  \centering
\includegraphics[width=0.6\textwidth]{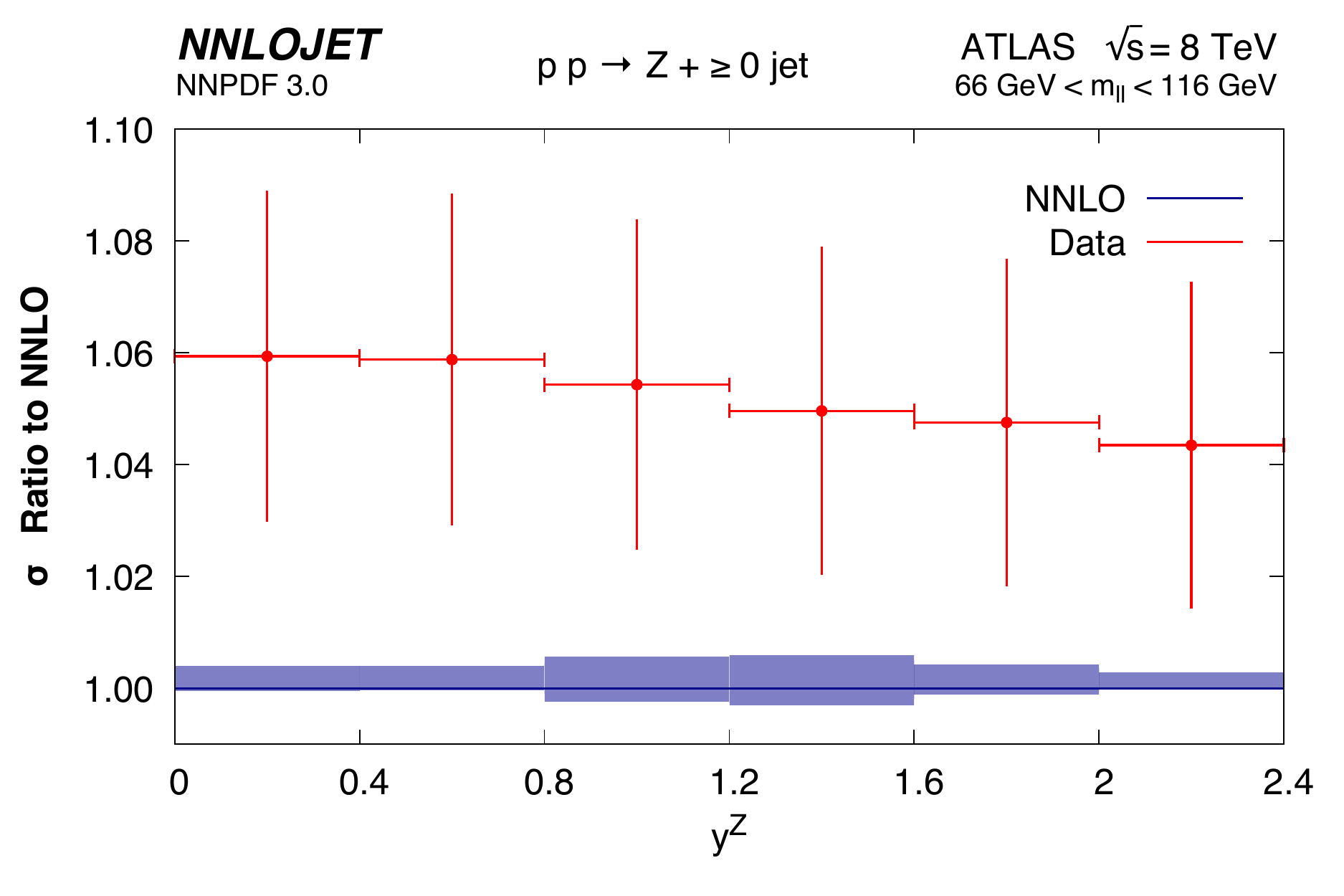}
\caption{The inclusive dilepton cross section for the same $|y^Z|$ bins as in Figure~\ref{fig:ATLASyz} and an invariant mass cut on final state leptons of $66$~GeV~$<m_{\ell\ell} < 116$~GeV. The ATLAS data is extracted from Ref.~\cite{ptzATLAS} by summing up the transverse momentum distributions in the respective bins. The vertical error bars are given by the luminosity uncertainty. The blue bands show the NNLO prediction with scale uncertainty. \label{fig:incATLASyz}}
\end{figure}
\begin{figure}
  \centering
\includegraphics[width=\textwidth]{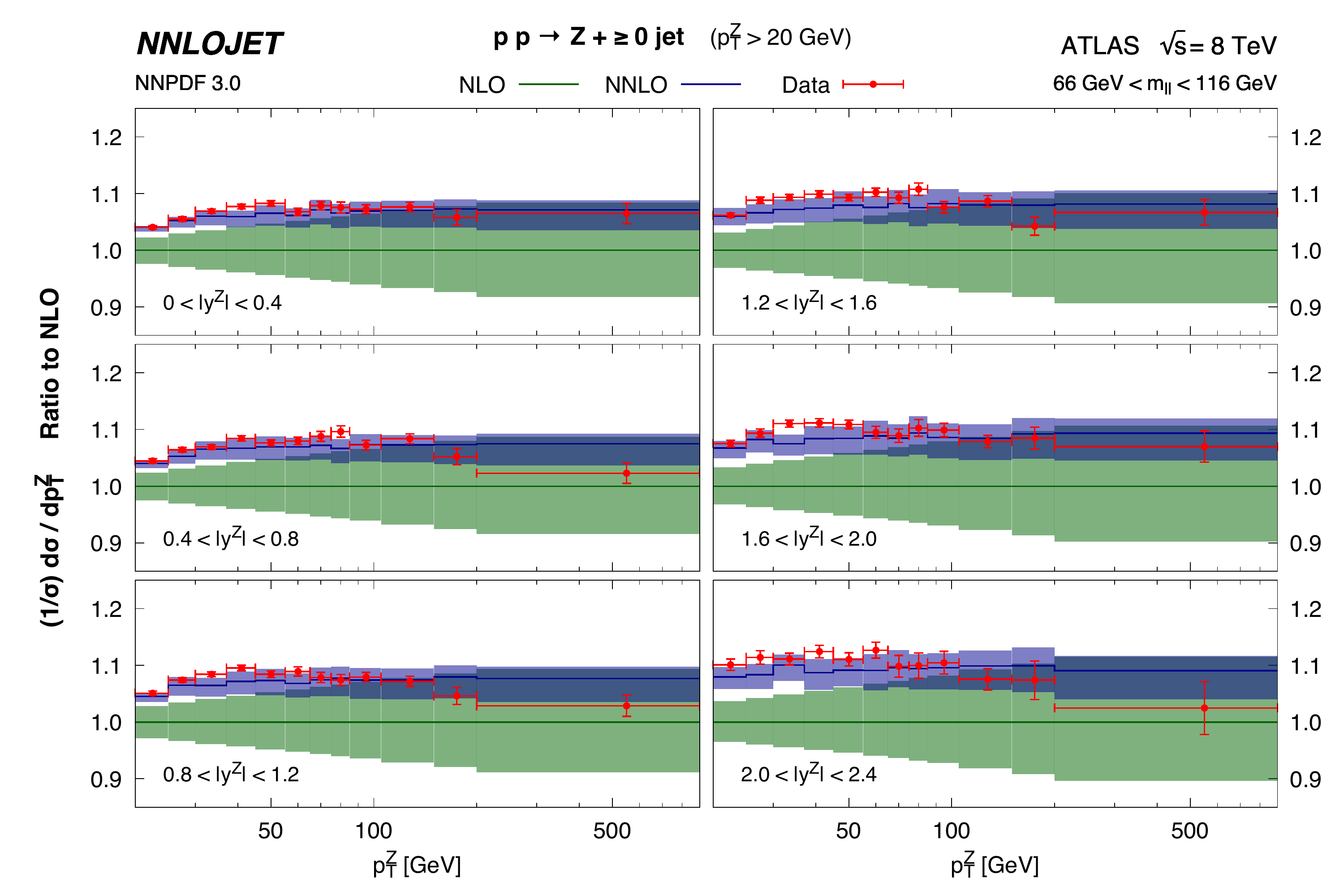}
\caption{The normalised double-differential transverse momentum distribution for the $Z$ boson in windows of rapidity of the $Z$ boson, $y^Z$, with an invariant mass cut on final state leptons of $66$~GeV~$<m_{\ell\ell} < 116$~GeV. The ATLAS data is taken from Ref.~\cite{ptzATLAS}.  The green bands denote the NLO prediction with scale uncertainty and the blue bands show the NNLO prediction.\label{fig:normATLASyz}}
\end{figure}

In order to eliminate the overall luminosity uncertainty, 
one again normalises the data to the 
inclusive fiducial dilepton cross section in the respective bins in $y^Z$. Figure~\ref{fig:incATLASyz} shows the ratio of ATLAS data and the NNLO cross section compared to the NLO prediction\footnote{Note that the ATLAS data is extracted from Ref.~\cite{ptzATLAS} by summing the bins of the $p_T^Z$ distributions in the various $y^Z$ bins. We have checked that this procedure applied to the different $m_{\ell\ell}$ bins shown in Figure~\ref{fig:incATLASmll} correctly reproduces the data.}.
As for the rapidity-integrated cross section
in this mass bin,
Figure~\ref{fig:incATLASmll}, we observe the NNLO prediction to be accurate to 0.9\%, and to underestimate the ATLAS 
data by about two standard deviations. 
The normalised double-differential distribution is shown in Figure~\ref{fig:normATLASyz}.   There is excellent agreement between the normalised NNLO prediction and the ATLAS data for all rapidity bins, again offering the possibility of further constraining the PDFs in this kinematic region.

\begin{figure}
  \centering
\includegraphics[width=\textwidth]{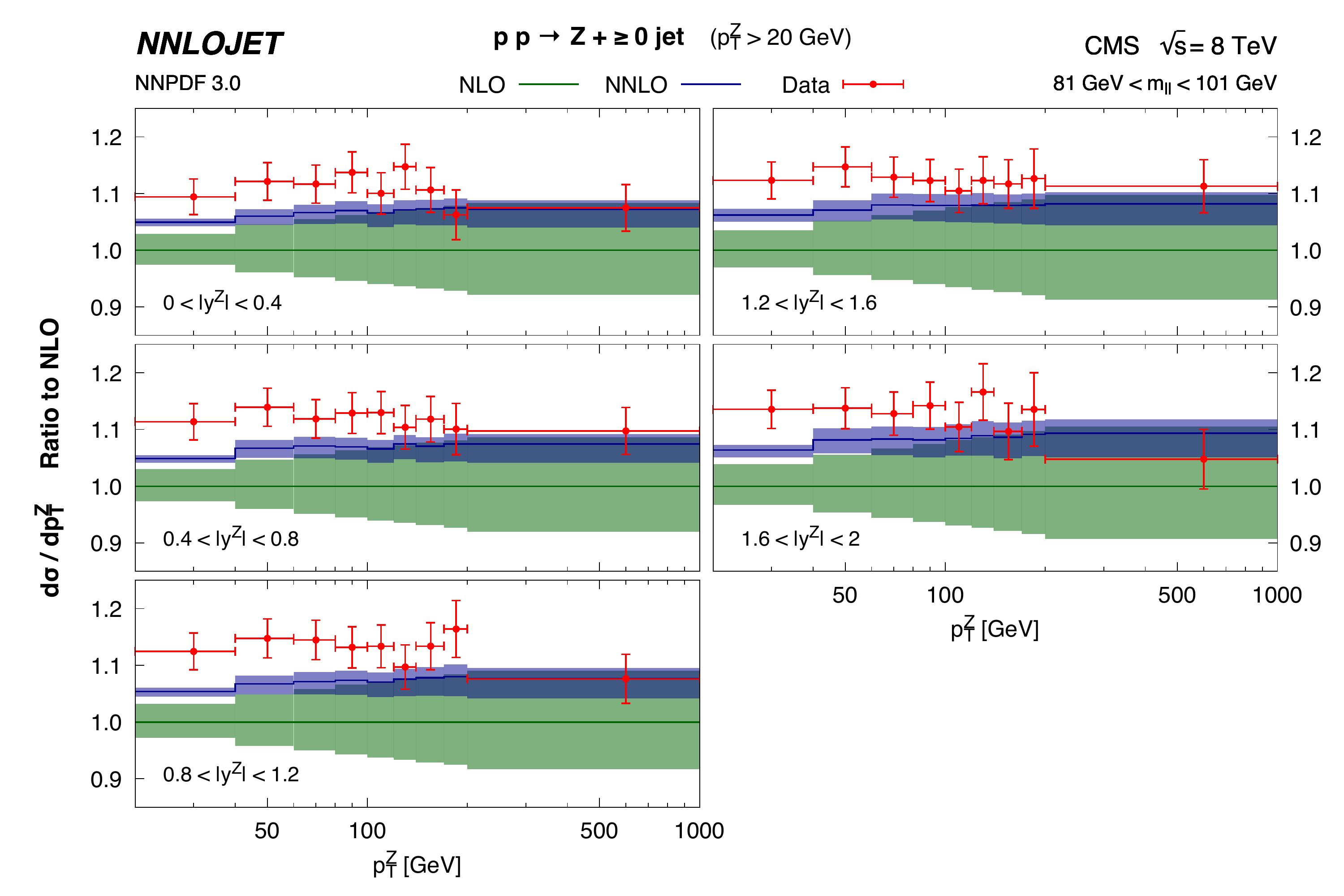}
\caption{The unnormalised double-differential transverse momentum distribution for the $Z$ boson in windows of rapidity of the $Z$ boson, $y^Z$, with an invariant mass cut on final state leptons of $81$~GeV~$<m_{\ell\ell} < 101$~GeV. The CMS data is taken from Ref.~\cite{ptzCMS}. The luminosity error is not shown. The green bands denote the NLO prediction with scale uncertainty and the blue bands show the NNLO prediction.\label{fig:CMSyz}}
\end{figure}
\begin{figure}
  \centering
\includegraphics[width=\textwidth]{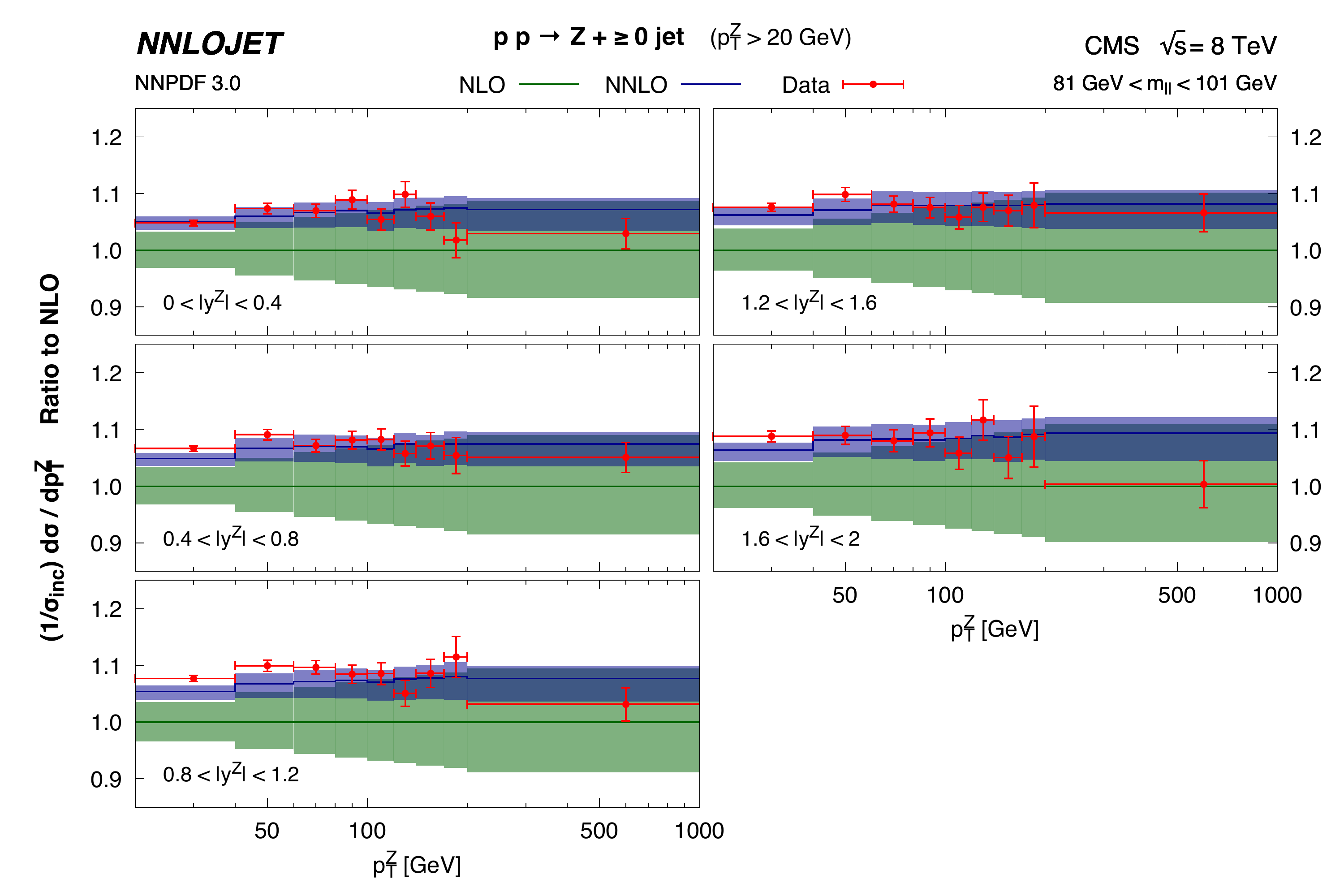}
\caption{The normalised double-differential transverse momentum distribution for the $Z$ boson in windows of rapidity of the $Z$ boson, $y^Z$, with an invariant mass cut on final state leptons of $81$~GeV~$<m_{\ell\ell} < 101$~GeV. The CMS data is taken from Ref.~\cite{ptzCMS}.  The green bands denote the NLO prediction with scale uncertainty and the blue bands show the NNLO prediction.\label{fig:normCMSyz}}
\end{figure}

The CMS measurement of the $Z$-boson transverse momentum distribution at 8~TeV~\cite{ptzCMS} concentrates on the
dilepton invariant mass range $81$~GeV~$<m_{\ell\ell} < 101$~GeV around the $Z$ resonance. The fiducial region
of this measurement, defined in Table \ref{tab:fiducial}, applies asymmetric cuts on the leading and sub-leading lepton, disregarding the lepton charge. 
To ensure the correct implementation of these cuts in \NNLOJET, and to maintain local cancellations between 
matrix elements and subtraction terms in the antenna subtraction method, 
 the lepton identification as leading and sub-leading 
had to be passed alongside the lepton charge through all subtraction terms and into the event analysis routines. 
The double-differential measurement is performed in $p_T^Z$ and $y^Z$. 
Comparing the absolute double-differential distributions of Figure~\ref{fig:CMSyz} to our NNLO predictions, we observe the same 
features as for the ATLAS measurement, with positive NNLO corrections at the level of 5--8\% and an NNLO scale uncertainty of 
2--4\%. Compared to NLO, inclusion of 
NNLO corrections brings the theoretical prediction closer to the experimental data, which are however still 
about 5-8\% larger than expected from theory.

In the CMS study~\cite{ptzCMS}, the normalised cross section is obtained by dividing the distributions by the fiducial 
cross section integrated over all bins in $y^Z$ (in contrast to ATLAS, which normalised to the fiducial cross sections
only integrated over the respective bin in $y^Z$). At NNLO, we obtain for the fiducial cross section with CMS cuts:
\begin{equation*}
\sigma_{\text{NNLO}}(81~\text{GeV} < m_{\ell\ell} < 101~\text{GeV}) =  450.6^{+2.7}_{-1.6} \text{ pb}.
\end{equation*}
The normalised distributions of CMS are compared to theory in Figure~\ref{fig:normCMSyz}, where excellent agreement 
is observed upon inclusion of the NNLO corrections. 

Compared to NNLO theory, both ATLAS and CMS fiducial cross section
measurements of the $Z$-boson transverse momentum
display a similar pattern of disagreement for the absolute distributions while having excellent agreement for the normalised distributions. 
The inclusion of the newly derived NNLO corrections to the transverse momentum distribution are crucial for a meaningful comparison between 
data and theory: (a) they reduce the theory uncertainty to a level that firmly establishes the discrepancy on 
the absolute cross sections, and (b) they modify the central value and the shape of the theory prediction to better agree 
with the data on the 
normalised distributions.

\section{Summary and Conclusions}
We have derived the NNLO QCD corrections to the production of $Z/\gamma^*$ bosons decaying to lepton pairs
at large transverse momentum, 
inclusive over the hadronic final state. 
 This observable has typically been very easy to measure but difficult to predict theoretically because of the necessity to be fully inclusive with respect to all QCD radiation.
 Our calculation is performed using the parton-level Monte Carlo generator \NNLOJET
which implements the antenna subtraction method for NNLO calculations of hadron collider observables. It extends our 
earlier calculation of $Z/\gamma^*$+jet production at this order~\cite{ZJNNLOus} and now also includes inclusive $Z/\gamma^*$ 
production.
We have performed a thorough comparison of theory predictions to the 8~TeV Run 1 data of the LHC
for cross sections defined over a fiducial region of lepton kinematics
 from the ATLAS~\cite{ptzATLAS} and CMS~\cite{ptzCMS} collaborations.
We observe that the NNLO corrections to the unnormalised distributions are moderate and positive, 
resulting in theory predictions that are closer to the experimental observations than at NLO. 
However, this improvement is much more dramatic when one considers distributions normalised to the relevant dilepton cross section.  This is true for both single and double-differential distributions.  While the normalised distributions for low $m_{\ell\ell}$ slices in comparison to ATLAS are dominated by the theoretical scale uncertainty on the inclusive dilepton cross section, for larger $m_{\ell\ell}$ and all $|y^Z|$ bins the agreement between NNLO theory and LHC data is excellent.  
With the reduction of the theoretical scale uncertainty 
at NNLO, a careful 
reinvestigation of the parametric ingredients to the NNLO theory predictions (parton distributions, strong coupling constant, electroweak parameters) 
appears now to be mandatory. Our calculation enables exactly these precision studies, and will allow a consistent 
inclusion of the precision data on the $Z$ transverse momentum distribution into NNLO determinations of parton 
distributions, leading to precision constraints at this order on the gluon distribution over a large 
momentum range. 

\acknowledgments
We would like to thank Daniel Froidevaux and Elzbieta Richter-Was for stimulating discussions concerning the ATLAS data, Dieter Zeppenfeld for enlightening comments that helped to shape the phenomenological study in this paper and Alexander M\"{u}ck for helpful comments concerning the impact of EW corrections to these observables.  
We also thank Xuan Chen, Juan Cruz-Martinez, James Currie and Jan Niehues for useful discussions and their many contributions to the \NNLOJET\ code.
This research was supported in part by the National Science Foundation under Grant NSF PHY11-25915,
in part by the Swiss National Science Foundation (SNF) under contracts 200020-162487 
and CRSII2-160814, in part by
the UK Science and Technology Facilities Council, in part by the Research Executive Agency (REA) of the European Union under the Grant Agreement PITN-GA-2012-316704  (``HiggsTools'') and the ERC Advanced Grant MC@NNLO (340983).

\end{document}